\documentclass[reprint,aps,prl,superscriptaddress,twocolumn,floatfix]{revtex4-2}
\usepackage{graphicx}
\usepackage{bm}
\usepackage[abs]{overpic}
\usepackage{amsmath,amssymb,amsfonts}
\usepackage{xcolor}

\usepackage[breaklinks=true, colorlinks,citecolor=blue,linkcolor=blue,urlcolor=blue]{hyperref}

\usepackage[abs]{overpic}
\usepackage{mathrsfs}

\newcommand{\be}{\begin{eqnarray}}
\newcommand{\ee}{\end{eqnarray}}
\newcommand{\nn}{\nonumber\\}

\renewcommand{\Im}{\operatorname{\mathfrak{Im}}}

\begin{document}

\title{Non-conservation of the valley density and its implications for the observation of the valley Hall effect}
\author{Alexander Kazantsev}
\email{alexander.kazantsev@manchester.ac.uk}
\affiliation{\noindent Department of Physics and Astronomy, University of Manchester, Manchester M13 9PL, UK}
\author{Amelia Mills}
\affiliation{\noindent Department of Physics and Astronomy, University of Manchester, Manchester M13 9PL, UK}
\author{Eoin O'Neill}
\affiliation{\noindent Department of Physics and Astronomy, University of Manchester, Manchester M13 9PL, UK}
\author{Hao Sun}
\affiliation{\noindent The Institute for Functional Intelligent Materials (I-FIM), 
National University of Singapore, 
4 Science Drive 2, Singapore 117544}
\author{Giovanni Vignale}
\affiliation{\noindent The Institute for Functional Intelligent Materials (I-FIM), 
National University of Singapore, 
4 Science Drive 2, Singapore 117544}
\author{Alessandro Principi}
\email{alessandro.principi@manchester.ac.uk}
\affiliation{\noindent Department of Physics and Astronomy, University of Manchester, Manchester M13 9PL, UK}

\begin{abstract}
We show that the conservation of the valley density in multi-valley insulators is broken in an unexpected way by the electric field that drives the valley Hall effect.  This implies that  time-reversal-invariant fully-gapped  insulators, in which no bulk or edge state crosses the Fermi level, can support a valley Hall current in the bulk and yet show no valley density accumulation  at the edges.  Thus, the valley Hall effect cannot be observed in such systems.
If the system is not fully gapped then valley density accumulation at the edges is possible. 
The accumulation has no contribution from undergap states and can be expressed as a Fermi surface average, for which we derive an explicit formula.  
We demonstrate the theory by calculating the valley density accumulations  in an archetypical valley-Hall insulator: a gapped graphene nanoribbon. Surprisingly, we discover that a net valley density polarization is dynamically generated for 
certain
edge terminations.
\end{abstract}

\maketitle

{\it Introduction}---The valley Hall effect (VHE) 
in non-topological systems has recently stirred considerable controversy~\cite{Gorbachev, ZhuEtAl, PhysRevLett.114.256601, PhysRevLett.99.236809, BeconciniPolini, SongVignale, RocheNikolic, Roche_2022, PhysRevB.103.115406, PhysRevB.103.115406}.  When the band structure features two valleys with a non-vanishing Berry curvature, electrons skew in the direction orthogonal to the applied electric field, even without a magnetic field. However, since the system is not topological, electrons from the two valleys skew in opposite directions
giving rise to a zero (charge) Hall current but a finite {\it valley} Hall current ${\bm j}_v({\bm r},t)$. This is defined as the difference between charge currents of electrons originating in opposite valleys.
When this current hits the edge of the system,
a valley density $n_v({\bm r},t)$ (or, more physically, orbital magnetization~\cite{PhysRevB.103.195309}), is expected to accumulate at its boundaries. 
This assumes 
that the valley density obeys a standard continuity equation~\cite{BeconciniPolini, SongVignale}. This seems a reasonable assumption: the two valleys are well-separated in momentum space, up to the point that they could ideally be taken as completely disconnected.

Some authors~\cite{Gorbachev,SongVignale} went further and claimed  that even a fully-gapped non-topological insulator 
such as graphene aligned with hexagonal boron nitride (hBN) \cite{PhysRevLett.114.256601, PhysRevLett.99.236809}
can exhibit nonlocal charge transport mediated by transverse  bulk undergap  valley currents.
The authors of Ref.~\cite{SongVignale} argued that, at finite temperature, the valley-density accumulation could drive a ``squeezed edge current" (parallel to the edges) in apparent agreement with experiment~\cite{ZhuEtAl}.  However, other authors~\cite{RocheNikolic, Roche_2022, PhysRevB.103.115406}  found from microscopic calculations no valley density accumulation or edge current in the simple graphene/hBN model. In the case of a {\it fully gapped insulator},  in which no bulk or edge state crosses the Fermi level, this leaves us with the following puzzle: on one hand, the electric field drives a finite dissipationless valley Hall current  in the bulk; on the other hand, time reversal symmetry implies that a valley density accumulation---a time-reversal-odd quantity---cannot appear in response to an electric field, unless there is dissipation, which is impossible with no states at the Fermi level. So where did the valley current go?

\begin{figure}[t]
\includegraphics[scale=0.29]{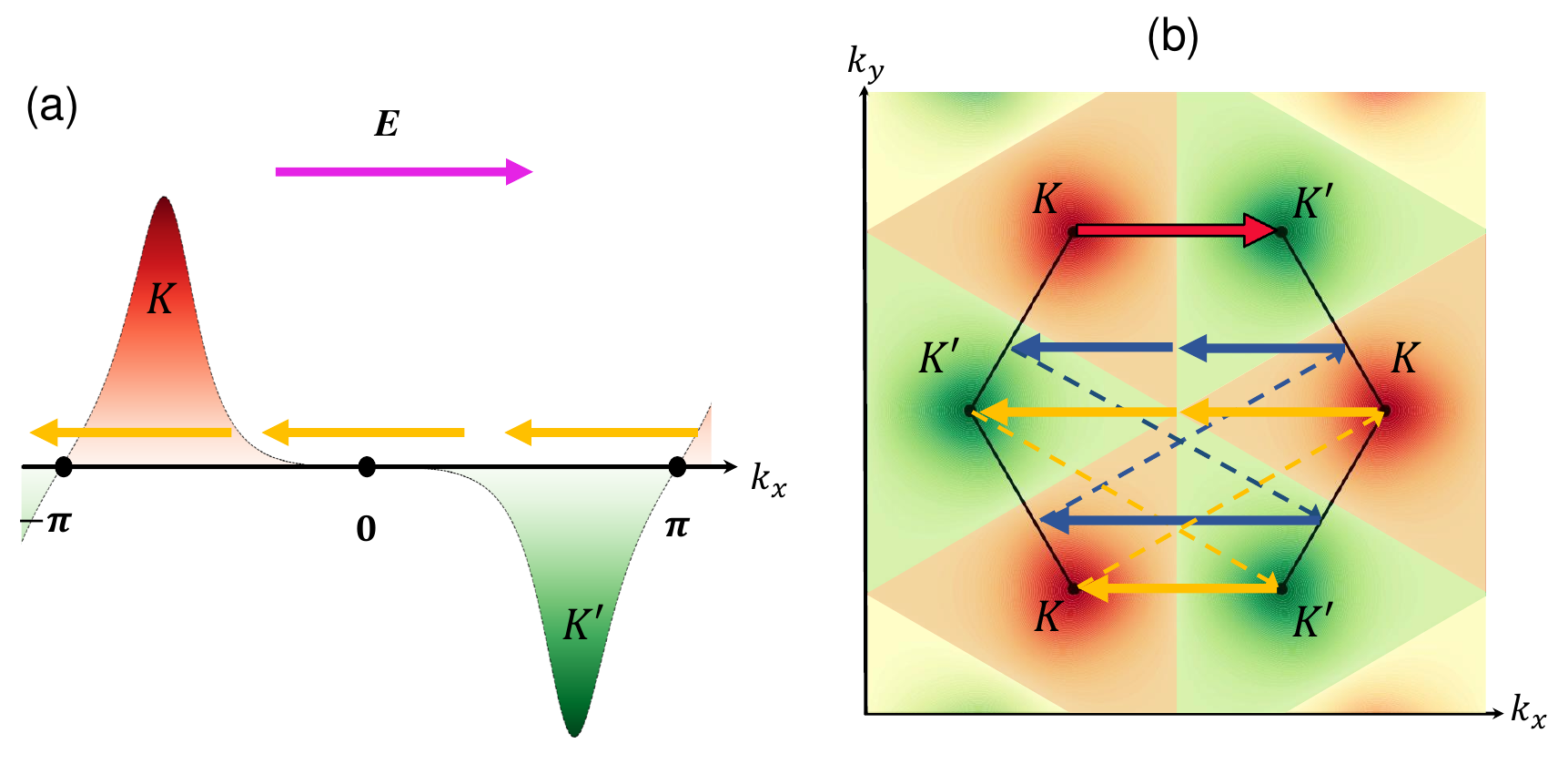}
\caption{\label{Fig1}
{Panel~(a): The cyclic flow of electrons (yellow arrows) in the one-dimensional Brillouin zone of a ribbon subject to electric field $E$ (pink arrow). The valley charge changes sign whenever the electron crosses the boundaries between the red and green regions ($k_x=0,\pm \pi$). 
Each electron performs half the cycle as a ``left-valley electron" and half as a ``right-valley electron". Also shown are the Berry curvature hot spots with a positive (negative) value near $K$ ($K'$). Due to opposite Berry curvatures 
in the two valleys, the result is a steady valley Hall current.  However, in a fully gapped insulator, at the end of the cycle each electron returns to its initial state,
thus no valley redistribution occurs. Panel (b): Two examples (blue and yellow arrows) of a similar flow in the two-dimensional Brillouin zone of the infinite system.}
}
\end{figure}

In this paper we solve the puzzle by observing that valley density does not satisfy a conventional continuity equation when an electric field
is present. 
The reason is that the electric field breaks the conservation of crystal momentum and 
therefore
of valley number, which depends explicitly on  it.
As a result, the bulk valley current is internally short-circuited as electrons flow from one valley to the other (and thus switch the sign of the Berry curvature) under the action of the very same electric field that drives the valley Hall current in the first instance.  This process is schematically 
shown
in Fig. \ref{Fig1}.

Our results imply that in a  (i) time-reversal invariant (ii) fully gapped insulator, as defined above, the undergap valley current cannot produce a valley density accumulation at the edge. This holds irrespective of presence/absence of spin-orbit coupling as long as the two conditions above are met and applies  equally well to systems based on graphene or transition metal dichalcogenides.  This makes observing the VHE impossible  in such systems unless, e.g., the valley degeneracy is lifted~\cite{Du2022, D0TC03712E, C9NR03315G} or carriers are selectively injected into a single valley~\cite{Mak_Pump}. 
In these cases, an electric (not valley) charge density is detected due to a non-zero net anomalous Hall effect. This is distinct from the VHE that was originally discussed in Refs.~\cite{Gorbachev, ZhuEtAl, PhysRevLett.114.256601} and for which the controversy exists.  Our result also implies that the non-local transport detected in Refs. \cite{Gorbachev, Sui2015, 10.1063/1.5094456, Shimazaki2015, Rebeca} must have been caused by partially occupied bulk or edge bands.

In metallic systems, which support a Fermi surface, our predictions are quite different from those of the conventional theory which assumes valley number conservation. In particular,  in our theory the accumulation depends on the form of the electronic wave functions near the edge. The length over which it occurs is not related to the carrier diffusion length as in, e.g., Ref.~\cite{BeconciniPolini}, but reflects the much shorter localization length of edge states, as  observed in some experiments \cite{Kawakami}, or the Fermi wavelength of bulk states.   Perhaps the most important result of this study is that valley density of equal sign can be generated on both edges simultaneously \footnote{Since valley number is closely related to orbital magnetic moment \cite{PhysRevB.103.195309} we expect similar results for orbital magnetization.}.

{\it Summary of main ideas}---We consider a generic system in the shape of a strip of finite width which is indefinitely extended along the $x$ axis. 
As we show below, the 
continuity equation satisfied by the valley density  is
\be \label{eq:anomalous_continuity_equation}
\partial_t n_v(y, t) + \partial_y  j^y_{v}(y, t) = -e^2 E(t) \sum_{k}S(k) \partial_{k}f_k(y)\,,
\ee
where the electric field  is in the $x$ direction, which is parallel to the edge, and the valley current is in the $y$ direction, perpendicular to the edge.  Edges are chosen to be parallel to one of the vectors connecting the two valleys in the Brillouin zone of the infinite system. One such vector (one of six) is shown in Fig.~\ref{Fig1}~(b) with a red arrow.  We call $k_x$ its direction in momentum space and $x$ in real space. In a strip, $k_x$ remains a good quantum number and serves as Bloch momentum in the one-dimensional Brillouin zone. For brevity, in what follows we will drop subscript ``$x$'' on $k_x$. The electronic states (in the absence of the electric field) have the form $\psi_{k,n}(x,y)=e^{ikx}u_{k,n}(x,y)/\sqrt{2\pi}$, where $n$ is the band index. 
The sum over $k$ in Eq.~(\ref{eq:anomalous_continuity_equation}) stands for  $\int dk/(2\pi)$.
The mixed electronic distribution $f_k(y) =   a^{-1}\int_0^a dx\sum_n  f_{k,n} \big| u_{k,n}(x, y) \big|^2$ is defined in terms of the electronic wave functions and the occupations of the corresponding states $f_{k,n}$, with the integral taken over one period $a$ in the $x$ direction. 
$S(k)$ is a ``valley charge" function (odd under time reversal), which is a smooth periodic function of $k$ in the Brillouin zone.  It assigns number $+1$ to states around one valley and $-1$ to states around the other valley.  The valley density operator 
is 
$\hat{n}_v (\bm{r})\equiv-(e/2)\sum_j \big\{S(\hat{k}_j), \delta(\bm{r}-\hat{\bm{r}}_j)\big\}$ where 
$\hat{\bm{r}}_j$ and $\hat{k}_j $ are the position and Bloch momentum operator (along the edge) of the $j$-th electron, respectively,
and $\{\hat A,\hat B\}\equiv\big(\hat A\hat B+\hat B\hat A\big)$.
The valley current density is 
$\bm{\hat{j}}_v (\bm{r})\equiv-(e/4)\sum_j \big\{S(\hat k_j), \big\{\hat{\bm{v}}_j\,, \delta(\bm{r}-\hat{\bm{r}}_j)\big\}\big\}$,  where $\hat{\bm{v}}_j$ is the velocity operator.
Since $\hat{k}$ is  conserved, 
$\hat{n}_v$ and $\bm{\hat{j}}_v (\bm{r})$
obey 
a conventional
continuity equation in the absence of the electric field. 
 
As shown below, in a fully gapped time-reversal invariant insulator, in which 
no edge or bulk state
crosses
the Fermi level, and at zero temperature, the right-hand side of Eq.~(\ref{eq:anomalous_continuity_equation}) completely cancels the contribution due to the current
on the left hand side. Thus,
the valley density accumulation vanishes, even though there is a finite valley current in the bulk.  
In all other cases the cancellation is not exact. The correct equation for the density accumulation rate in the absence of relaxation processes is then
$\partial_t n_v(y, t) =  -Q_s(y)$, where
 the source term 
\be \label{eq:valley_density_source}
Q_s(y)=\frac{e^2E}{a}\int_0^a dx \sum_{k, n} \left(\partial_k f_{k,n}\right)S(k)\big|u_{k,n}(x,y)\big|^2,
\ee
is a Fermi surface property.
Note that $Q_s(y)$ cannot be written, in general, as the divergence of a current. In fact, this is only possible if 
its integral across the strip vanishes, which implies that density accumulates at one edge and depletes at the other~\footnote{A similar situation was discussed in Ref.~\cite{Shi_prl_2006} for the spin current, with the difference that there the non-conservation of spin density was caused by an intrinsic spin-orbit torque, while here it is caused by the very same electric field that drives the valley Hall effect.}.  However, 
if the width of the strip is macroscopically large, 
the source term is localized on the edges. 
One can then define the ``effective current" $I_s$,
obtained by integrating Eq.~(\ref{eq:valley_density_source}) across a given edge,
that feeds 
the valley number accumulation thereat. It can be split as
$I_s=I_s^e+I_s^b$, where 
$I_{s}^e=e^2E\sum_{k, e} \left(\partial_k f_{k,e}\right)S(k)$  is the contribution of the edge states. Here, the sum over $e$ is that over the edge states.
The 
contribution of the bulk states, $I_{s}^b$,
can be obtained in terms of the probability amplitude for 
the propagating Bloch waves to scatter off the edge [see Eq.~(\ref{eq:bulk_something}) below]. Once $I_s$ is known, the  valley number accumulation can be estimated as $I_s\tau_{tr}$, where $\tau_{tr}$ is the intra- or inter-valley momentum relaxation time for the bulk or edge states' contribution, respectively.

{\it Anomalous continuity  equation}---We consider a 2D crystal periodic in the $x$ direction with  period $a=1$ and with the edges positioned at $y=0$ and $y=-W$.  
A uniform electric field of magnitude $E$ oscillating at frequency $\omega$ is applied along the $x$ direction. 
For conciseness, hereafter we set $\hbar=1$.
Thus the conductance quantum $e^2/h$ equals $e^2/(2\pi)$, where $e$ is the electron charge. 
From the Kubo formula~\cite{doi:10.1143/JPSJ.12.570, giuliani_vignale_2005}, the $y$ component of the valley current (averaged over $x$) is~\footnote{See the supplemental online material for more details.}
\be\label{eq:current_response}
j_v^y(y, \omega) &=& iEe^2\sum_{k, n,n'} \int_0^y dy' 
(\varepsilon_{k,n}-\varepsilon_{k,n'})
\nn
&\times&
S(k){\cal L}_{k,nn'}(\omega) {\cal W}_{k,nn'}(y')\mathcal{A}_{k,n'n},
\ee
and the valley density (also averaged over $x$) 
\be\label{eq:density_resp}
&& n_v(y,\omega) = -\frac{iEe^2}{\omega+i0}\sum_{k,n}
(\partial_k f_{k,n})\, S(k)\, {\cal W}_{k,nn}(y)
\nn
&& -Ee^2\sum_{k,n,n'}
S(k)
{\cal L}_{k,nn'} (\omega){\cal W}_{k,nn'}(y)\mathcal{A}_{k,n'n},
\ee
where ${\cal L}_{k,nn'}(\omega) \equiv (f_{k,n}-f_{k,n'})/(\omega+\varepsilon_{k,n}-\varepsilon_{k,n'}+i0)$ is the usual Lindhard factor~\cite{giuliani_vignale_2005}, 
${\cal W}_{k,nn'}(y) \equiv \int_0^1 dx \,u_{k,n}^\dag(x,y)u_{k,n'}(x,y)$, and ${\cal A}_{k,n'n}= \int_0^1 dx\int_{-W}^0 dy\, u_{k,n'}^\dag(x,y) i\partial_k u_{k,n}(x,y)$ is the Berry connection.
The Fourier transform of Eq.~\eqref{eq:anomalous_continuity_equation} follows directly~\cite{Note3} from Eqs.~\eqref{eq:current_response} and \eqref{eq:density_resp}:
\be 
\label{eq:discontinuity_eq}
-i\omega n_v(y,\omega) + \partial_y j_v^y(y, \omega)
=-e^2E
\sum_k
S(k)
\partial_k f_{k}(y).
\ee

{\it The vanishing of valley density accumulation}---Let us first assume that the system is a time-reversal invariant fully gapped insulator,  such that no bulk or edge state crosses the Fermi level. The first term on the right hand side of Eq.~(\ref{eq:density_resp})  vanishes because $\partial_k f _{k,n}=0$, since 
no band
crosses the Fermi level. 
Due to time-reversal symmetry, the second line on the right hand side of Eq.~(\ref{eq:density_resp}) is proportional to  $\omega$~\cite{Note3}, so  the valley density accumulation vanishes 
 for a
static electric field. 
This result implies that  $\partial_y  j_{v}^y(y)$ can be different from zero---as it must necessarily be, since the valley Hall current is finite in the bulk but vanishes at the edges---yet this finite divergence does not cause any density change at the edge or anywhere else. The resolution of this paradox is provided by the anomalous term on the right hand side of Eq.~(\ref{eq:anomalous_continuity_equation})  which exactly matches the divergence term on the left-hand side when the system is  fully gapped.

{\it The source of valley density}---Let us now consider the case  in which the system is not fully gapped and some states cross the Fermi level.  Then the cancellation between the anomalous term and divergence of the current is not perfect. Indeed,
the first term on the right hand side of Eq.~(\ref{eq:density_resp})  causes the density to grow at a constant rate, leading to a breakdown of linear response theory unless a limiting momentum relaxation mechanism
is taken into account. The Fermi surface term, obtained by multiplying Eq.~\eqref{eq:density_resp} by $-i\omega$ and taking the $\omega \to 0$ limit, is the ``source term" 
$Q_s(y)$
in Eq.~(\ref{eq:valley_density_source}).  It receives contributions from both bulk and edge states, both decaying away from the edge, the latter exponentially and the former oscillating at half the Fermi wavelength. As discussed above,
the integral of $Q_s(y)$ over $y$ across a single edge can be interpreted as an effective current $I_s$  that feeds the density accumulation thereat.  The contribution of bulk states to $I_s$ is (at $y=0$)
\be\label{eq:bulk_something}
I_{s}^b=-2 e^2E\sum_{\lambda,k,p>0} \partial_{k}f^\lambda_{k,p}
\Im \bigg[\frac{ [v^\lambda_{k,p}]^\dag v^\lambda_{k,-p}R_\lambda(k, p)}{p+i0}\bigg],
\ee
where momentum integration is restricted to the valley with valley number $+1$, $p$ is momentum in the $y$ direction measured from the valley bottom, $v^\lambda_{k,p}$ are envelope amplitudes of propagating stationary states, labelled by index $\lambda$,  $R_\lambda(k,p)$ is the reflection probability amplitude ($|R_{\lambda}(k,p)|=1$) (see \cite{Note3} for details).

\begin{figure}[t]
\includegraphics[scale=0.83]{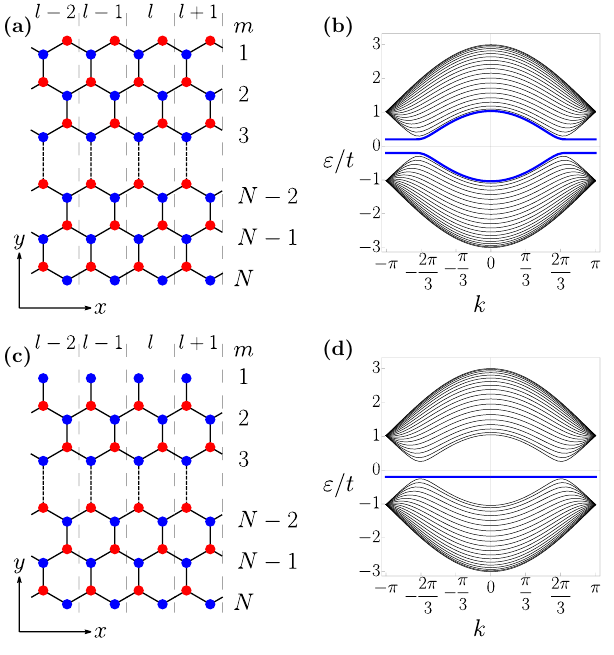}
\caption{\label{fig:nanoribbon} 
Panel (a) and (c):
Gapped graphene nanoribbon with two zig-zag edges 
and a zig-zag and a bearded edge (at the top), respectively.
Red (blue) discs signify atoms of the $A$ ($B$) sublattice. 
Panel (b) and (d):
Band structures of nanoribbons of panels (a) and (c), respectively, for $N=20$ and $\Delta=0.2t$. 
}
\end{figure} 

{\it Example: ``gapped graphene"}---To illustrate the general theory developed above, we calculate the valley Hall current and valley density accumulation rate for  a nanoribbon of ``gapped graphene"---a model system that captures some aspects of  monolayer graphene on a gap-inducing hBN substrate. For the nanoribbon we consider two terminations: a) zig-zag boundaries on both edges [Fig.~\ref{fig:nanoribbon}(a)] and 
b) a zig-zag and a bearded edge [Fig.~\ref{fig:nanoribbon}(c)]. Each unit cell, labelled by an integer $l$,  contains $N$ horizontal rows,  each labelled by an integer $m$. Each row contains two atoms of sublattices $A$ and $B$ as shown in Fig.~\ref{fig:nanoribbon}(a), except the edge rows, where one atom may be missing as shown in Fig.~\ref{fig:nanoribbon}(c).  The two sublattices, $A$ and $B$,  have different on-site potentials $\pm \Delta$. Electrons are assumed to hop only between nearest neighbors. We neglect spin-orbit interaction and therefore consider spinless electrons. The $y$ coordinate will take integer values to indicate the row and half-integer values to mark the position halfway between the rows.  

The band structures for the two terminations, shown in Figs.~\ref{fig:nanoribbon}(b) and~(d), respectively, feature two bulk bands separated by a gap equal to $2|\Delta|$  with minima at  $k=\pm 2\pi/3$. These points define the two valleys in the one-dimensional Brillouin zone. The blue lines show bands of edge states.

\begin{figure}[t]
\includegraphics[scale=0.85]{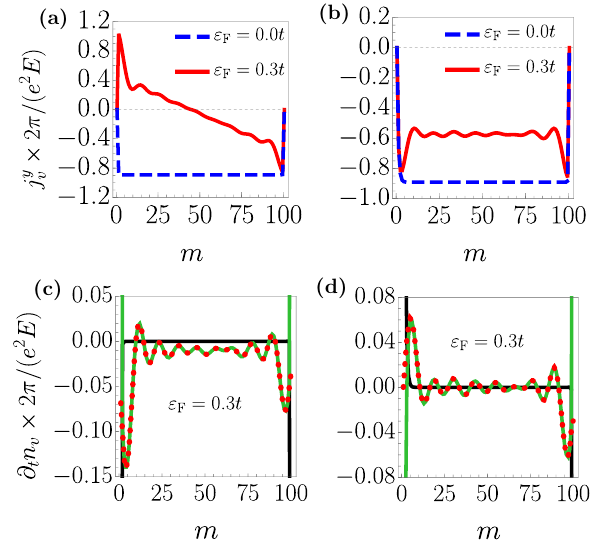}
\caption{\label{fig:current_distr}
Panel (a): Gapped graphene nanoribbon with zig-zag edges: the blue dashed (red solid) line shows the valley Hall current as a function of position at Fermi energy $\varepsilon_{\rm F}=0$  
($\varepsilon_{\rm F}=0.3t$).  
Panel (b): Same as in (a) for a nanoribbon with one zig-zag and one bearded edge. 
Panel (c): 
Contribution to the valley density accumulation rate from the valley Hall current (green), from the non-conservation term (black), and their sum, {\it i.e.,} the total accumulation rate (red dotted), in the nanoribbon with zig-zag edges. Panel (d):  Same as in (c) for the nanoribbon with one zig-zag and one bearded edge. In all plots 
$N=100$,  $\Delta=0.1t$. In plots (c) and (d) $\varepsilon_{\rm F}=0.3t$.
}
\end{figure}

Our main results are presented in Fig.~\ref{fig:current_distr}. For a Fermi energy in the gap ($\varepsilon_{\rm F}=0$) and at zero temperature, 
we find that $n_v(m, 0)=0$ for either termination, consistent with the absence of states at the Fermi level.  At the same time $j^y_v(m+1/2, 0)=-Ee^2\cdot\mbox{sign}(\Delta)/(2\pi)+O(\Delta/t)$ for $1\leq m\leq N-1$ as shown in panels (a) and (b), blue line: this is the undergap current associated with the nearly-quantized Hall conductance  (the actual value $-0.9$ deviates from the ideal quantized value $-1$ due to the finite bandwidth of the model)~\cite{Note3}.

When the system is doped, the current distributions differ dramatically for the two terminations, as shown by the red lines in Figs.~\ref{fig:current_distr}~(a) and~(b).  In the case of the double zig-zag termination the current shows a linear variation across the ribbon [red line in (a)], changing sign about the center of the ribbon.  This is contrary to our intuition, which would suggest an approximately constant current in the bulk.  
Of greater physical interest, however, is the valley density accumulation rate shown in Fig.~\ref{fig:current_distr}~(c). There is a significant cancellation between $-\partial_y j_v^y$ (green line) and the non-conservation term (black line) at the edges. Their sum results in a density accumulation rate displaying oscillations (red dots) on the scale of half the Fermi wavelength and two spikes of equal signs at the edges.  The fact that the accumulation rate does not integrate to zero is the result of the anomaly on the right-hand side of Eq. \eqref{eq:anomalous_continuity_equation}: valley number is pumped from one valley into the other via a partially filled band of edge states connecting the two [upper blue line in Fig. \ref{fig:nanoribbon}~(b)].  This opens an intriguing possibility of generating a net valley density polarization by purely electrical means, as opposed to the standard optical methods. Note, however,  that the form of valley density accumulation rate cannot be predicted from the valley Hall current alone and depends on the boundary conditions. Because valley number and orbital magnetic moment are closely related \cite{PhysRevB.103.195309}, the same effect should emerge for orbital magnetization, as was indeed found in Ref. \cite{PhysRevB.104.165403}.

The zig-zag+bearded termination presents us with a more familiar scenario. Fig.~\ref{fig:current_distr} (b) shows that the valley Hall current is approximately constant   ($-0.55$ in units of $e^2E/(2\pi)$ at $\varepsilon_{\rm F}=0.3t$) in the bulk.   At the same time the valley density accumulation rate [panel (d)] has spikes of opposite signs on the two edges (red dots).  In this case, valley density is transported from one edge to the other. The reason for the overall valley number conservation is, in contrast to the previous example, absence of partially filled bands connecting the two valleys.     

{\it Conclusion}---The modified continuity equation~(\ref{eq:anomalous_continuity_equation})  allows us to explain how a non-vanishing undergap valley current can coexist with a vanishing valley density accumulation in a fully gapped non-topological time-reversal-invariant system with perfectly degenerate valleys.  Any valley density accumulation requires the existence of states at the Fermi level and furthermore it is a  dissipative process which requires a scattering mechanism to reach a steady state. We have provided closed expressions for calculating valley density accumulation rates  on the edges of a two-dimensional material and we have applied them to the gapped graphene model: these formulas show that the connection between bulk currents and measurable edge accumulations is much more complex than previously suspected. This, in particular, leads us to surmise that any physical system in which evidence of the VHE has been found either by  Kerr rotation microscopy~\cite{Lee2016}  or by non-local resistance measurements \cite{Gorbachev, Sui2015, 10.1063/1.5094456, Shimazaki2015, Rebeca} cannot be a true insulator but must have partially populated bulk or edge states.

{\it Acknowledgments}---A.P. acknowledges support from the European Commission under the EU Horizon 2020 MSCA-RISE-2019 programme (project 873028 HYDROTRONICS). A.P. and A.K. acknowledge support from the Leverhulme Trust under the grant RPG-2019-363.  H.S. and  G.V. were supported by the Ministry of Education, Singapore, under its Research Centre of Excellence award to the Institute for Functional Intelligent Materials (I-FIM, project No. EDUNC-33-18-279-V12).

\end{document}


\title{Supplemental Material}

\begin{abstract}
\end{abstract}

\maketitle

Everywhere below $\hbar=1$.

\section{The setup}\label{sec:setup}

We consider a 2D system periodic in the $x$ direction and of finite width $W$ in the $y$ direction. We will take the period in the $x$ direction to be equal to one. We divide the system into unit cells, labelled by index $l$. We assume that the periodic direction is chosen such that the valley index remains a good quantum number, {\it i.e.,} edges are chosen to be parallel to the separation between the valleys in momentum space (one of the edges will be taken to be at $y=0$ and the other at $y=-W$). The Bloch eigenstates are denoted as $\psi_{k,n}(x,y) = u_{k,n}(x,y)e^{ikx}/\sqrt{2\pi}$, where $k$ is the wavevector in the $x$ direction and $n$ is the band index. To take account of the spin-orbit interaction, we will assume that both $\psi_{k,n}$ and $u_{k,n}$ are two-dimensional spinors.
We apply the electric field of magnitude $E$ in the direction parallel to the edges ({\it i.e.,} the $x$ direction), {\it i.e.,} the Hamiltonian is perturbed by a potential term $e E {\hat x}$, where $-e$ is the electron charge and ${\hat x}$ the position operator. 

For graphene nanoribbon with zig-zag and bearded edges (which preserve the valley number), the role of coordinates $(x,y)$ is played by unit cell number $l$ and combined index  $(m,\sigma)$, where $m=1\dots N$ denotes the two-atom horizontal row in each unit cell and $\sigma$ the atom ($A$ or $B$) within it (see Fig.~1~(a), (c) in the main text). Note that a row may miss an atom of either sublattice, as in Fig.~1 (c), where it misses an $A$ atom in row $1$ of each unit cell. Everywhere below when displaying the results for densities and currents in the nanoribbon, the position on the $y$ axis will only be resolved down to the row number $m$. Therefore, the $y$ coordinate will be replaced by a discrete index that will take integer values marking the position in a certain row and half-integer values marking the position (half-way) between the rows. 

\section{Position operator in Bloch state basis}
The derivation of the position operator representation in the Bloch state basis here follows closely Ref.~\cite{LL}.  Consider a wavepacket $\chi(x,y)$ that is a superposition of Bloch eigenstates with coefficients $\chi_{k,n}$ 
\be \label{eq:chi_FT_def}
\chi(x,y)&=&\sum_{n}\int dk \chi_{k,n}\psi_{k,n}(x,y).
\ee
The application of the position operator $\hat{x}$ to $\chi(x,y)$ gives
\be
\hat{x}\chi(x,y)&=&\sum_{n}\int dk\, \chi_{k,n}\,x\psi_{k,n}(x,y)\nonumber\\&=&\sum_{n}\int dk \chi_{k,n}\Big(-i\partial_k \psi_{k,n}(x,y)\nonumber\\&&\qquad\quad{}+ie^{ikx}\partial_k u_{k,n}(x,y)/\sqrt{2\pi}\Big).
\ee
 We integrate by parts in the first term, while discarding the boundary contribution. Furthermore, we expand $i\partial_k u_{k,n}$ in the series of functions $u_{k,n}$ so that
\be \label{eq:Berry_connection_intro}
i\partial_{k}u_{k,n}(x,y)=\sum_{n'} u_{k,n'}(x,y){\cal A}_{k,n'n},
\ee
 with ${\cal A}_{k,n'n}=\int_0^1 dx\int_{-W}^0 dy\, u^\dag_{k,n'}(x,y)i\partial_k u_{k,n}(x,y)$. As a result, we obtain (lightening the notation by suppressing $x$ and $y$ dependence)
\be
\hat{x}\chi&=&\sum_n\int dk \Big[(i\partial_k \chi_{k,n})\psi_{k,n}+\chi_{k,n}{\cal A}_{k,n'n}\psi_{k,n'} \Big]\nonumber\\&=&\sum_{nn'}\int dk dk'\;\psi_{k',n'}\;x_{k'n';kn}\;\chi_{k,n},\
\ee
 where
\be \label{eq:x_representation_bloch}
x_{k'n';kn}=-i\partial_{k}\delta(k-k')\delta_{nn'}+\delta(k-k'){\cal A}_{k,n'n}.
\ee

\section{Valley number and valley Hall current operators}
Here everything refers to just one electron, the generalization to many electrons is trivial. The operator that gives valley density at position $\bm{x}=(x, y)$ has the form 
\be\label{eq:density_op}
\hat{n}_v(x,y)=-\frac{e}{2}\big\{\hat{n}(x,y), S(\hat{k})\big\},
\ee
where $\hat{n}(x,y)=|x,y\rangle\langle x,y|$ is the particle density operator at position $(x,y)$ (with any spin polarization) and the curly brackets stand for an anticommutator. 

Note that the particle density operator satisfies the continuity equation
\be\label{eq:cont_eq}
\partial_t \hat{n}(x,y,t)+\bm{\nabla}\hat{\bm{j}}(x,y,t)=0,
\ee
 with a particle number current
\be
\hat{\bm{j}}(x,y)=\frac{1}{2}\big\{\hat{\bm{v}}, n(x,y)\big\},
\ee
 where $\hat{\bm{v}}$ is the velocity operator. Operator $S(\hat{k})$, on the other hand, is a constant of motion. Therefore, taking the time derivative of Eq. \eqref{eq:density_op} at time $t$, we obtain
\be
\partial_t \hat{n}_v(x,y,t)+\bm{\nabla}\hat{\bm{j}}_v(x,y,t)=0,
\ee
where
\be\label{eq:def_of_curr}
\hat{\bm{j}}_v(x,y)&=&-\frac{e}{2}\big\{\hat{\bm{j}}(x,y), S(\hat{k})\big\}\nonumber\\&=&-\frac{e}{4}\big\{\{\hat{\bm{v}}, \hat{n}(x,y)\}, S(\hat{k})\big\}.
\ee
Note also that $\hat{n}(x,y)=\delta(\bm{x}-\hat{\bm{r}})$, where $\hat{\bm{r}}$ is the electron's position operator.

\section{Kubo formula for valley number and valley Hall current}

The Kubo formula \cite{doi:10.1143/JPSJ.12.570, giuliani_vignale_2005} allows one to calculate the change in the ensemble average of an observable caused by a perturbation to first order in its strength. If the Hamiltonian is perturbed by an operator $\hat{B}\exp(-i\omega t)$ then, long after the perturbation was turned on, the ensemble average of the observable $\hat{A}$ will be oscillating at the frequency $\omega$ with the amplitude
\be \label{eq:gen_Kubo}
A(\omega)=\sum_{\alpha\beta}\frac{f_{\alpha}-f_{\beta}}{\omega+\varepsilon_\alpha-\varepsilon_\beta+i0}A_{\alpha\beta}B_{\beta\alpha}.
\ee
Here, $\alpha$ and $\beta$ label eigenstates of the unperturbed Hamiltonian, whose eigenvalues are $\varepsilon_\alpha$ and $\varepsilon_\beta$, respectively. Furthermore, $A_{\alpha\beta}$ and $B_{\beta\alpha}$ are matrix elements of operators $\hat{A}$ and $\hat{B}$ between eigenstates $\alpha$ and $\beta$, while $f_{\alpha,\beta}=(\exp[(\varepsilon_{\alpha,\beta}-\varepsilon_{\rm F})/(k_B T)]+1)^{-1}$ are the occupation numbers of such states.

For the nanoribbon, the unperturbed stationary states are labelled by the value of (quasi)momentum $k$ and by the band index $n$.  In the case of a harmonic electric field applied parallel to the edge,  $\hat{B}=eE\hat{x}$. Using Eq.~(\ref{eq:x_representation_bloch}), the matrix elements of the perturbation are then given by the equation
\be 
B_{k'n';kn}&=&eE x_{k'n';kn}\nonumber\\
&=&eE\big(-i\partial_{k}\delta(k-k')\delta_{nn'}+\delta(k-k')\mathcal{A}_{k,n'n}\big).\nonumber\\
\ee
 The role of observable $\hat{A}$ is played by either $\hat{n}_v(x,y)$ or $\hat{\bm{j}}_v(x, y)$, depending on the response function under consideration. Matrix elements of operator $\hat{n}_v(x,y)$ between stationary states are
\be\label{eq:en_vee}
[\hat{n}_v(x,y)]_{kn;k'n'}&=&-\frac{e}{2}\langle k,n|\big\{\hat{n}(x,y), S(\hat{k})\big\}|k',n'\rangle\nonumber\\&=&-\frac{e}{2}\big(S(k)+S(k')\big)\langle k,n|\hat{n}(x,y)|k',n'\rangle\nonumber\\
&=&-\frac{e}{4\pi}\big(S(k)+S(k')\big)e^{-i(k-k')x}\nonumber\\&&\quad\quad\quad{}\times u_{k,n}^\dag(x,y)u_{k',n'}(x,y),
\ee
 where we have used the fact that, by definition, $\hat{n}(x,y)=|x,y\rangle\langle x,y|$ and $\langle x,y|k,n\rangle=\psi_{k,n}(x,y)$. For $k=k'$, Eq.~\eqref{eq:en_vee} becomes
\be\label{eq:en_vee_simple}
[\hat{n}_{v}(x,y)]_{kn;kn'}=-\frac{e}{2\pi} u^\dag_{k,n}(x,y)u_{k,n'}(x,y) S(k).
\ee
Matrix elements of operator $\hat{\bm{j}}_v(x,y)$ are given by
\be
&&[\hat{\bm{j}}_v(x,y)]_{kn;k'n'}=-\frac{e}{2}\big(S(k)+S(k')\big)[\hat{\bm{j}}(x,y)]_{kn;k'n'}.\nonumber\\
\ee
 At $k'=k$ we obtain
\be\label{eq:two_curr}
&&[\hat{\bm{j}}_v(x,y)]_{kn;kn'}=-eS(k)[\hat{\bm{j}}(x,y)]_{kn;kn'}.
\ee
 Note that due to Eq. \eqref{eq:cont_eq} the following identity is true
\be
\bm{\nabla}\langle k,n|\hat{\bm{j}}(x,y)|k,n'\rangle &=&-i\langle k,n|[\hat{H}, \hat{n}(x,y)]|k,n'\rangle\nonumber\\
&=&i(\varepsilon_{k,n'}-\varepsilon_{k,n})\langle k,n|\hat{n}(x,y)|k,n'\rangle.\nonumber\\
\ee
 Plugging into this equation $\hat{n}(x,y)=|x,y\rangle\langle x,y|$ and $\psi_{k,n}(x,y)=u_{k,n}(x,y)/\sqrt{2\pi}$ will give
\be
&&\bm{\nabla}\big[\hat{\bm{j}}(x,y)\big]_{kn,kn'}=\frac{i(\varepsilon_{k,n'}-\varepsilon_{k,n})}{2\pi}u_{k,n}^\dag(x,y)u_{k,n'}(x,y).\nonumber\\
\ee
 Let us integrate this equation along the length of one unit cell and using the periodicity of $[\bm{j}(x,y)]_{kn,kn'}$ in the $x$ direction discard the boundary terms. The result will be
\be
&&\partial_y\int_0^1 dx\,\big[\hat{j}^y(x,y)\big]_{kn,kn'}\nonumber\\&&\qquad\qquad{}=\frac{i(\varepsilon_{k,n'}-\varepsilon_{k,n})}{2\pi}\int_0^1 dx\,u_{k,n}^\dag(x,y)u_{k,n'}(x,y).\nonumber\\
\ee
 Now using the fact that  $\hat{j}^y$ must vanish at the boundary of the strip at $y=0$, we can integrate this equation with respect to $y$ to obtain
\be\label{eq:auxil}
&&\int_0^1 dx\,\big[\hat{j}^y(x,y)\big]_{kn,kn'}=\frac{i(\varepsilon_{k,n'}-\varepsilon_{k,n})}{2\pi}\int_0^y dy'\int_0^1 dx'\nonumber\\&&\qquad\qquad\qquad\qquad\qquad\qquad{}\times\,u_{k,n}^\dag(x',y')u_{k,n'}(x',y').\nonumber\\
\ee
 From Eqs. \eqref{eq:two_curr} and \eqref{eq:auxil} we can now obtain that
\begin{widetext}
\be\label{eq:final_matrix_el}
\int_0^1 dx\,\big[\hat{j}_v^y(x,y)\big]_{kn,kn'}=-\frac{ieS(k)}{2\pi}\big(\varepsilon_{k,n'}-\varepsilon_{k,n}\big)\int_0^y dy'\int_0^1 dx'\,u_{k,n}^\dag(x',y')u_{k,n'}(x',y').
\ee
 In calculating the linear response, we will average over the length of the unit cell in the $x$ direction, so that Eq. \eqref{eq:final_matrix_el} will turn out to be useful. The linear response formula for the valley Hall current, averaged over the length of the unit cell, has the form
\be\label{eq:curr}
 j_v^y(y, \omega)&=&eE \int_0^1 dx \sum_{nn'}\int dk dk'\frac{(f_{k,n}-f_{k',n'})}{\omega+\varepsilon_{k,n}-\varepsilon_{k',n'}+i0}[j^y_v(x,y)]_{kn;k'n'}(-i\partial_k\delta(k-k')\delta_{nn'}+\delta(k-k'){\cal A}_{k,n'n}).
\ee 
 Integrating by parts the first term in the round brackets and then evaluating the integral with respect to $k'$ will give
\be\label{eq:matrix_element}
&&j_v^y(y,\omega)=\frac{ieE}{\omega+i0}\int_0^1 dx \sum_{n}\int dk\, \partial_k f_{k,n} \, [j^y_v(x,y)]_{kn;kn}\nonumber\\&&\qquad\qquad\qquad\qquad\qquad\qquad\qquad\qquad{}+eE\int_0^1 dx\sum_{nn'}\int dk \frac{f_{k,n}-f_{k,n'}}{\omega+\varepsilon_{k,n}-\varepsilon_{k,n'}+i0}[j^y_v(x,y)]_{kn;kn'}{\cal A}_{k,n'n}.
\ee
 The first term vanishes because the unit cell average  $\int_0^1 dx \,[j_v^y(x,y)]_{kn,kn}$ in that term vanishes, see Eq.~\eqref{eq:final_matrix_el} at $n=n'$.  Plugging Eq. \eqref{eq:final_matrix_el} into the second line, we obtain
\be\label{eq:l_a_current}
 j^y_v(y, \omega)=ie^2E \int_0^y dy' \sum_{nn'}\int \frac{dk}{2\pi}\frac{(f_{k,n}-f_{k,n'})(\varepsilon_{k,n}-\varepsilon_{k,n'})}{\omega+\varepsilon_{k,n}-\varepsilon_{k,n'}+i0}\int_0^1 dx'\, u_{k,n}^\dag(x',y')u_{k,n'}(x',y')S(k){\cal A}_{k,n'n},
\ee
 which gives Eq. (3) of the main text. Next, let us calculate the valley number response. According to Eq.~\eqref{eq:gen_Kubo}, it is given by the equation, averaged across the length of the unit cell,
\be
 n_{v}(y,\omega)=eE\int_0^1 dx\sum_{nn'}\int dk dk'\frac{f_{k,n}-f_{k',n'}}{\omega+\varepsilon_{k,n}-\varepsilon_{k',n'}+i0}[n_v(x,y)]_{kn,k'n'}(-i\partial_k\delta(k-k')\delta_{nn'}+\delta(k-k'){\cal A}_{k,n'n}).
\ee
 Integrating by parts the first term in the round brackets and using Eq.~\eqref{eq:en_vee_simple}, we obtain 
\be \label{eq:delta_n_vee}
 n_v(y,\omega)&=&-\frac{ie^2E}{\omega+i0}\sum_{n}\int \frac{dk}{2\pi}\partial_k f_{k,n} S(k)\int_0^1 dx\,u_{k,n}^\dag(x,y)u_{k,n}(x,y)\nonumber\\&&{}\qquad\qquad\qquad-e^2E\sum_{nn'}\int\frac{dk}{2\pi}\frac{f_{k,n}-f_{k,n'}}{\omega+\varepsilon_{k,n}-\varepsilon_{k,n'}+i0}S(k)\int_0^1 dx\, u_{k,n}^\dag(x,y)u_{k,n'}(x,y){\cal A}_{k,n'n},
\ee
\end{widetext}
 which is Eq. (4) in the main text. In the limit of zero frequency, if there is a partially filled band such that $\partial_k f_{k,n}$ does not vanish, the first term gives the dominant contribution. One can demonstrate that if the system is gapped and time-reversal symmetry is not broken, the second term on the right-hand side of Eq.~\eqref{eq:delta_n_vee} is of order $O(\omega)$ and, therefore, vanishes.  This fact, which is demonstrated in Sect.~\ref{sect:time_reversal} below, was used in the main text to discard its contribution to the valley density accumulation.

\section{The valley density response is gauge invariant}
Let us demonstrate in this section that the valley density response, Eq. (4) in the main text, is gauge invariant. The first line of Eq. (4) is obviously gauge invariant so let us turn to the second line. This is proportional to 
\be\label{eq:sec_line}
&&P(\omega+i0)=\sum_{n,n'}\int \frac{dk}{2\pi}S(k)\frac{f_{k,n}-f_{k,n'}}{\omega+\varepsilon_{k,n}-\varepsilon_{k,n'}+i0}\nonumber\\&&\quad\quad\quad\quad\quad{}\times \int_0^1 dx\, u_{k,n}^\dag(x,y)u_{k,n'}(x,y){\cal A}_{k,n'n}.
\ee

Let us change phases of all the stationary states as follows
\be\label{eq:modified_states}
u_{k,n}(x,y)\to e^{-i\phi_{k,n}}u_{k,n}(x,y).
\ee
 After this the Berry connection ${\cal A}_{k, n'n}$ changes as 
\begin{widetext}
\be
&&{\cal A}_{k,n'n}\to  \int_{0}^1dx\int_{-W}^0 dy\, u_{k,n'}^\dag(x,y)\Big(i\partial_k u_{k,n}(x,y)\Big)e^{i(\phi_{k,n'}-\phi_{k,n})}\nonumber\\&&\qquad\qquad\qquad\qquad\qquad\qquad{}+ \int_0^1 dx\int_{-W}^0 dy\, u_{k,n'}^\dag(x,y)u_{k,n}(x,y)(\partial_k\phi_{k,n})e^{i(\phi_{k,n'}-\phi_{k,n})}.
\ee
 Using the normalization condition this can be rewritten as
\be
{\cal A}_{k,n'n}\to {\cal A}_{k,n'n}e^{i(\phi_{k,n'}-\phi_{k,n})}+\delta_{n,n'}(\partial_k \phi_{k,n}).
\ee
 Let us now plug the modified states \eqref{eq:modified_states} into Eq. \eqref{eq:sec_line}. This will result in the following change
\be
P(\omega+i0)&\to&\sum_{n,n'}\int \frac{dk}{2\pi}S(k)\frac{f_{k,n}-f_{k,n'}}{\omega+\varepsilon_{k,n}-\varepsilon_{k,n'}+i0} \int_0^1 dx\,u_{k,n}^\dag(x,y)u_{k,n'}(x,y)e^{i(\phi_{k,n}-\phi_{k,n'})}\nonumber\\
&&
\quad\quad\quad
\quad\quad\quad
\quad\quad\quad
\quad\quad\quad
\quad\quad\quad
\times{}\Big({\cal A}_{k,n'n}e^{i(\phi_{k,n'}-\phi_{k,n})}+\delta_{n,n'}(\partial_k \phi_{k,n})\Big).
\ee
 The second term in the round brackets multiplies $f_{k,n}-f_{k,n'}$ and disappears. The phases in the rest of the equation cancel each other. Therefore, $P(\omega+i0)$ will not change, which means that it is gauge invariant.
\end{widetext}

\section{Proof of the continuity equation}

In this section we demonstrate the validity of Eq. (1) in the main text.   Let us multiply Eq. \eqref{eq:delta_n_vee} by $-i\omega$. By representing $\omega$ as $\omega=\omega+\varepsilon_{k,n}-\varepsilon_{k,n'}-(\varepsilon_{k,n}-\varepsilon_{k,n'})$, we can rewrite the result as
\begin{widetext}
\be\label{eq:a_long_one}
-i\omega  n_{v}(y,\omega)&=&{}-Ee^2\sum_{n}\int\frac{dk}{2\pi}\partial_k f_{k,n}S(k)\int_0^1 dx\, u_{k,n}^\dag(x,y)u_{k,n}(x,y)\nonumber\\
&&{}+iEe^2\sum_{nn'}\int\frac{dk}{2\pi}(f_{k,n}-f_{k,n'})S(k)\int_0^1 dx\, u_{k,n}^\dag(x,y)u_{k,n'}(x,y){\cal A}_{k,n'n}\nonumber\\&&{}-iEe^2\sum_{nn'}\int\frac{dk}{2\pi}\frac{(f_{k,n}-f_{k,n'})(\varepsilon_{k,n}-\varepsilon_{k,n'})}{\omega+\varepsilon_{k,n}-\varepsilon_{k,n'}+i0}S(k)\int_0^1 dx\, u_{k,n}^\dag(x,y) u_{k,n'}(x,y){\cal A}_{k,n'n}.
\ee
 Let us work on the sum in the second line of this equation, 
\be\label{eq:aye}
I(y)\equiv \sum_{nn'}\int\frac{dk}{2\pi}(f_{k,n}-f_{k,n'})S(k)\int_0^1 dx\, u_{k,n}^\dag(x,y)u_{k,n'}(x,y){\cal A}_{k,n'n}.
\ee
 Let us divide this sum in two across the minus sign and relabel $n$ to $n'$ and vica versa in the second sum. The result reads
\be 
I(y)=\sum_{nn'}\int\frac{dk}{2\pi}f_{k,n}S(k)\int_0^1 dx\, u_{k,n}^\dag(x,y)u_{k,n'}(x,y){\cal A}_{k,n'n}-\sum_{nn'}\int\frac{dk}{2\pi}f_{k,n}S(k)\int_0^1 dx\, u_{k,n'}^\dag(x,y)u_{k,n}(x,y){\cal A}_{k,nn'}.\nonumber\\
\ee
 Now let us plug into this equation the definition for ${\cal A}_{k,nn'}$
\be
{\cal A}_{k,nn'}\equiv \int_0^1 dx'\int_{-W}^0 dy'\, u_{k,n}^\dag(x',y')\Big(i\partial_k u_{k,n'}(x',y')\Big)=-\int_0^1 dx'\int_{-W}^0 dy' \Big(i\partial_k u_{k,n}^\dag(x',y')\Big)u_{k,n'}(x',y').
\ee
 The result will read
\be
I(y)&=&\sum_{nn'}\int_{0}^1 dx\int_0^1 dx'\int_{-W}^0 dy'\int\frac{dk}{2\pi}f_{k,n}S(k)u_{k,n}^\dag(x,y)u_{k,n'}(x,y) u_{k,n'}^\dag(x',y')\Big(i\partial_k u_{k,n}(x',y')\Big)\nonumber\\{}&&+\sum_{nn'}\int_0^1 dx\int_0^1 dx' \int_{-W}^0 dy'\int\frac{dk}{2\pi}f_{k,n}S(k)u_{k,n'}^\dag(x,y)u_{k,n}(x,y)\Big(i\partial_k u_{k,n}^\dag(x',y')\Big)u_{k,n'}(x',y').
\ee
Let us first perform summation over $n'$ by using the completeness relation
\be
\sum_{n'}u_{k,n'}(x,y)u_{k,n'}^\dag(x',y')=1_2\sum_{l}\delta(y-y')\delta(x-x'-l),
\ee
 where $1_{2}$ is the identity operator in the spin space, and then perform integration over $x'$ and $y'$. This will lead to
\be
I(y)=\sum_{n}\int \frac{dk}{2\pi}f_{k,n}S(k)\int_0^1 dx\,u_{k,n}^\dag(x,y)\Big(i\partial_k u_{k,n}(x,y)\Big)+\sum_{n}\int\frac{dk}{2\pi}f_{k,n}S(k)\int_0^1 dx\Big(i\partial_k u_{k,n}^\dag(x,y)\Big)u_{k,n}(x,y).\nonumber\\
\ee
 Joining the two sums together we obtain
\be\label{eq:aye_final}
I(y)=\sum_{n}\int_0^1 dx\int\frac{dk}{2\pi}f_{k,n}S(k)i\partial_k\Big(u_{k,n}^\dag(x,y)u_{k,n}(x,y)\Big).
\ee

Plugging this into Eq. \eqref{eq:a_long_one}, we obtain
\be 
-i\omega  n_v(y,\omega)=&-&Ee^2\sum_n\int\frac{dk}{2\pi}\partial_k f_{k,n}S(k)\int_0^1 dx\, u_{k,n}^\dag(x,y)u_{k,n}(x,y)\nonumber\\&-&Ee^2\sum_{n}\int\frac{dk}{2\pi}f_{k,n}S(k)\int_0^1 dx\, \partial_k\Big(u_{k,n}^\dag(x,y)u_{k,n}(x,y)\Big)\nonumber\\
&-&iEe^2\sum_{n,n'}\int \frac{dk}{2\pi}\frac{(f_{k,n}-f_{k,n'})(\varepsilon_{k,n}-\varepsilon_{k,n'})}{\omega+\varepsilon_{k,n}-\varepsilon_{k,n'}+i0}S(k)\int_0^1 dx \,u_{k,n}^\dag(x,y)u_{k,n'}(x,y){\cal A}_{k,n'n}.
\ee
 We combine the first two lines to obtain
\be\label{eq:fir}
-i\omega n_{v}(y,\omega)=&-&Ee^2\sum_{n}\int\frac{dk}{2\pi}\int_0^1 dx\,\partial_k\Big(f_{k,n}u_{k,n}^\dag(x,y)u_{k,n}(x,y)\Big)S(k)\nonumber\\
&-&iEe^2\sum_{n,n'}\int \frac{dk}{2\pi}\frac{(f_{k,n}-f_{k,n'})(\varepsilon_{k,n}-\varepsilon_{k,n'})}{\omega+\varepsilon_{k,n}-\varepsilon_{k,n'}+i0}S(k)\int_0^1 dx\, u_{k,n}^\dag(x,y)u_{k,n'}(x,y){\cal A}_{k,n'n}.
\ee

From Eq.~\eqref{eq:l_a_current}, the divergence of the current is obtained to be
\be\label{eq:sec}
\partial_y j_v^y(y,\omega)=iEe^2\sum_{nn'}\int\frac{dk}{2\pi}\frac{(f_{k,n}-f_{k,n'})(\varepsilon_{k,n}-\varepsilon_{k,n'})}{\omega+\varepsilon_{k,n}-\varepsilon_{k,n'}+i0}S(k)\int_0^1 dx\,u_{k,n}^\dag(x,y)u_{k,n'}(x,y){\cal A}_{k,n'n},
\ee
which is the term on the second line of Eq.~(\ref{eq:fir}).
Combining Eq. \eqref{eq:fir} and \eqref{eq:sec}, we obtain Eq. (1) in the main text.
\end{widetext}

\section{Lightning speed derivation of the continuity equation}

The continuity equation derived in the previous section can be easily derived using Heisenberg equations of motion. Everywhere below operators are given in the Heisenberg picture. The equation of motion for the valley density has the form
\be
\partial_t \hat{n}_v(\bm{r}, t)=i\big[\hat{H}(t)+eE(t)\hat{x}(t), \hat{n}_v(\bm{r},t)\big].
\ee
 Recall that in the absence of the electric field valley density satisfies a continuity equation, {\it i.e.},
\be
i\big[\hat{H}(t), \hat{n}_v(\bm{r},t) \big]=-\bm{\nabla}\hat{\bm{j}}_v(\bm{r},t),
\ee
 with the current defined as in Eq. \eqref{eq:def_of_curr}. Now using definition \eqref{eq:density_op} and the fact that 
\be
[\hat{x}, S(\hat{k})]=iS'(\hat{k}),
\ee
 where the prime indicates a derivative, one can obtain
\be
[\hat{x}(t), \hat{n}_v(\bm{r},t)\big]=-\frac{ie}{2}\big\{S'\big(\hat{k}(t)\big), \hat{n}(\bm{r},t)\big\}.
\ee
 Thus the equation of motion for the valley density has the form
\be
\partial_t \hat{n}_v(\bm{r},t)+\bm{\nabla}\hat{\bm{j}}_v(\bm{r},t)=\frac{e^2E(t)}{2}\big\{S'\big(\hat{k}(t)\big), \hat{n}(\bm{r},t)\big\}.\nonumber\\
\ee
 Replacing all the operators in this equation with their many-body versions and then calculating the ensemble average to first order in $E$ will give Eq. (1) in the main text.

\section{Time reversal symmetry implies no static valley polarization}
\label{sect:time_reversal}
Consider Eq. (4) in the main text and take the second line from that equation. It is proportional to 
\begin{widetext}
\be \label{eq:p}
P(\omega+i0)=\sum_{n,n'}\int \frac{dk}{2\pi}S(k)\frac{f_{k,n}-f_{k,n'}}{\omega+\varepsilon_{k,n}-\varepsilon_{k,n'}+i0}\int_0^1 dx \,u_{k,n}^\dag(x,y)u_{k,n'}(x,y){\cal A}_{k,n'n}.
\ee
 Assume presence of time-reversal symmetry which acts on the electron's wavefunction as
\be
T\psi(x,y)=-i\sigma_y \psi^*(x,y).
\ee
 For the Bloch states this implies that
\be
(-i\sigma_y)u_{k,n}^*(x,y)=\alpha_{k,n}u_{-k,\tilde{n}}(x,y),
\ee
 or
\be\label{eq:transform}
u_{k,n}^*(x,y)=\alpha_{k,n}i\sigma_y u_{-k,\tilde{n}}(x,y),
\ee
 where $\tilde{n}$ labels another stationary state which satisfies $\varepsilon_{k,n}=\varepsilon_{-k,\tilde{n}}$ and $\alpha_{k,n}$ is a possible phase factor.  Plugging Eq.~\eqref{eq:transform} into the connection ${\cal A}_{k,n'n}$ in Eq. \eqref{eq:p}, we obtain
\be
&&{\cal A}_{k,n'n}=i\int_0^1 dx\int_{-W}^0 dy \,u^t_{-k,\tilde{n}'}(x,y)\Big(\partial_k u^*_{-k,\tilde{n}}(x,y)\Big)\alpha_{k,n'}\alpha_{k,n}^*\nonumber\\&&\qquad\qquad\qquad\qquad\qquad\qquad{}+i\int_0^1dx\int_{-W}^0 dy\,\alpha_{k,n'}\Big(\partial_k\alpha_{k,n}^*\Big)u^t_{-k,\tilde{n}'}(x, y)u^*_{-k,\tilde{n}}(x,y),
\ee
 where superscript $t$ stands for `transposed'. Interchanging the electron wavefunctions in the bilinear products and using the normalization condition, this equation can  be rewritten as
\be\label{eq:conn_trans}
{\cal A}_{k,n'n}&=&i\int_0^1 dx\int_{-W}^0 dy\, u^\dag_{-k,\tilde{n}}(x,y)\Big(-\partial_{k}u_{-k,\tilde{n}'}(x,y)\Big)\alpha_{k,n'}\alpha_{k,n}^*+i\alpha_{k,n}\Big(\partial_k \alpha_{k,n}^*\Big)\delta_{n,n'}\nonumber\\&=&{\cal A}_{-k,\tilde{n}\tilde{n}'}\alpha_{k,n'}\alpha^*_{k,n}+i\alpha_{k,n}\Big(\partial_k \alpha_{k,n}^*\Big)\delta_{n,n'}.
\ee

Analogously, 
\be\label{eq:prod_trans}
u^\dag_{k,n}(x,y)u_{k,n'}(x,y)=\alpha_{k,n}\alpha_{k,n'}^*u_{-k,\tilde{n}'}^\dag(x,y)u_{-k,\tilde{n}}(x,y).
\ee

Let us now substitute Eqs. \eqref{eq:conn_trans} and \eqref{eq:prod_trans} into Eq. \eqref{eq:p}. We obtain
\be
P(\omega+i0)&=&\sum_{n,n'}\int\frac{dk}{2\pi}S(k)\frac{f_{k,n}-f_{k,n'}}{\omega+\varepsilon_{k,n}-\varepsilon_{k,n'}+i0}\int_0^1 dx \,u_{-k,\tilde{n}'}^\dag(x,y)u_{-k,\tilde{n}}(x,y)\alpha_{k,n}\alpha_{k,n'}^*\nonumber\\&&\quad\quad\quad\quad\quad\quad{}\times\Big({\cal A}_{-k, \tilde{n}\tilde{n}'}\alpha_{k,n'}\alpha_{k,n}^*+i\alpha_{k,n}\Big(\partial_k \alpha_{k,n}^*\Big)\delta_{n,n'}\Big).
\ee
 The term in the round brackets proportional to $\delta_{n,n'}$ gives a vanishing contribution to the sum and overall all the phases disappear.  Let us also take into account that $\varepsilon_{k,n}=\varepsilon_{-k,\tilde{n}}$, $\varepsilon_{k,n'}=\varepsilon_{-k,\tilde{n}'}$ and, consequently, $f_{k,n}=f_{-k, \tilde{n}}$ and $f_{k,n'}=f_{-k, \tilde{n}'}$. Let us make use of these identities to make the summand only depend on $\tilde{n}$ and $\tilde{n}'$ and then also note that summation over $n$ and $n'$ is the same as summation over $\tilde{n}$ and $\tilde{n}'$. Changing the summation variables from the former to the latter and then relabelling these back to $n$ and $n'$, we obtain
\be
P(\omega+i0)=\sum_{n,n'}\int\frac{dk}{2\pi}S(k)\frac{f_{-k,n}-f_{-k,n'}}{\omega+\varepsilon_{-k,n}-\varepsilon_{-k,n'}+i0}\int_0^1 dx\,u_{-k,n'}^\dag(x,y)u_{-k,n}(x,y){\cal A}_{-k, nn'}.
\ee
 Changing the integration variable from $k$ to $-k$, relabelling the summation indices from $n$ to $n'$ and vice versa, and using the fact that $S(-k)=-S(k)$, we obtain
\be\label{eq:conclus}
P(\omega+i0)=-\sum_{n,n'}\int\frac{dk}{2\pi}S(k)\frac{f_{k,n}-f_{k,n'}}{-\omega+\varepsilon_{k,n}-\varepsilon_{k,n'}-i0}\int_0^1 dx\,u_{k,n}^\dag(x,y)u_{k,n'}(x,y){\cal A}_{k,n'n}=-P(-\omega-i0).
\ee
 If the system is fully gapped, at small frequency the $i0$ prescription is irrelevant and can be neglected so that  Eq.~\eqref{eq:conclus} implies that $P(\omega)=-P(-\omega)$, which means that the second line of Eq. (4) is of order $O(\omega)$ and can be neglected at zero frequency.
\end{widetext}

\section{Valley Hall current at vanishing frequency}
At vanishing frequency, the following relation holds (compare Eqs. \eqref{eq:l_a_current} and \eqref{eq:aye}):
\be
 j_v^y(y, 0)=ie^2E\int_{0}^y dy'\,I(y').
\ee
Plugging the result \eqref{eq:aye_final} into this equation, we obtain
\be\label{eq:zero_freq_curr}
&&j_v^y(y, 0)=-e^2E \int_0^y dy'\int_0^1 dx' \sum_{n}\int\frac{dk}{2\pi}f_{k,n}S(k)\nonumber\\&&\qquad\quad\qquad\qquad{}\times\partial_{k}\Big(u_{k,n}^\dag(x',y')u_{k,n}(x',y')\Big).
\ee

\section{Effective valley current}

Let us try to evaluate the integral
\be\label{eq:Q}
&&Q_s(y)=e^2E\sum_n\int\frac{dk}{2\pi}\Big(\partial_{k}f_{k, n}\Big)S(k)\nonumber\\&&\qquad\qquad\qquad\quad{}\times\int_0^1 dx\,u^\dag_{k,n}(x,y)u_{k,n}(x,y)
\ee
 in the thermodynamic limit, {\it i.e.,} for the ribbon width $W\to \infty$. Assume that the chemical potential is in the conduction or valence band. Due to the factor $\partial_k f_{k,n}$ the integral over $k$ and sum over $n$ are restricted to the Fermi surface  which consists of two disjoints parts, one in each valley. Due to time reversal symmetry the two parts give equal contributions to the integral, so we can calculate just one and multiply the result by two, {\it i.e.,}
\be\label{eq:new_Q}
&&Q_s(y)=2e^2E\sum_n\int\limits_{S(k)=1}\frac{dk}{2\pi}\Big(\partial_{k}f_{k, n}\Big)\nonumber\\&&\qquad\qquad\qquad\quad{}\times\int_0^1 dx\,u^\dag_{k,n}(x,y)u_{k,n}(x,y).
\ee
 Everything from now on refers to the valley with valley number $+1$. Assuming that in the sum over $n$ and integral over $k$ we never go too far away from the bottom of the valley, we can use the envelope wave function description for the stationary states. Suppose energy eigenstates for a system without boundaries are described by a set of multicomponent valence (conduction) band envelope amplitudes $v^\lambda_{k,p}$ with energies $\varepsilon_{\lambda}(k,p)$, where $p$ is the component of momentum along $y$ measured from the bottom of the valley and $\lambda$ is a discrete label counting the stationary states. Now let us introduce the boundaries at $y=0$ and $y=-W$. Assume that the scattering off the boundaries does not mix  eigenstates with different values of $\lambda$ and that $\varepsilon_\lambda(k,-p)=\varepsilon_\lambda(k,+p)$. Then the  valence (conduction) band envelope wave-functions for the system with boundaries will have the form
\be\label{eq:state}
&&u^\lambda_{k,m}(y)={\cal N}_\lambda(k, p_m)\Big(v^\lambda_{k, p_m}e^{ip_my}\nonumber\\&&\qquad\qquad\qquad{}+R_\lambda(k, p_m)v^\lambda_{k,-p_m}e^{-ip_m y}\Big),
\ee  
 which is nothing but a sum of an incident and a reflected wave. Here $R_\lambda(k,p)$ is the probability amplitude for scattering off the boundary at $y=0$, momenta $p_m$ take discrete positive values  (labelled by $m$), with the distance between them approaching $\pi/W$ as $W\to\infty$, and an overall factor ${\cal N}_\lambda(k,p_m)$ is chosen such that
\be
\int_{-W}^0 dy \, [u_{k, m}^{\lambda}]^\dag(y) u_{k,m}^\lambda(y)=1.
\ee
 Note that at large $W$
\be\label{eq:norm}
|{\cal N}_\lambda(k,p)|^2=\frac{1}{2W}+O\bigg(\frac{1}{W^2}\bigg).
\ee
 In terms of the envelope wave-functions the expression for $Q_s(y)$ takes the form
\be\label{eq:new_new_Q}
&&Q_s(y)\nonumber\\&&\quad{}=2e^2E\sum_{m,\lambda}\int\limits_{S(k)=1}\frac{dk}{2\pi}\Big(\partial_{k}f^\lambda_{k, m}\Big) [u^\lambda_{k,m}]^\dag(y)u^\lambda_{k,m}(y),\nonumber\\
\ee
 Plugging Eq. \eqref{eq:state} into this equation, we obtain
\be\label{eq:new_new_new_Q}
&&Q_s(y)=4e^2E\sum_{m,\lambda}\int\limits_{S(k)=1}\frac{dk}{2\pi}\Big(\partial_{k}f^\lambda_{k, m}\Big)\nonumber\\&&\quad{}\times\bigg(1+\Re\Big([v^\lambda_{k,p_m}]^\dag v^\lambda_{k,-p_m}R_\lambda(k,p_m)e^{-2ip_m y}\Big)\bigg),\nonumber\\
\ee
 where we used the normalization condition  $|v_{k,p}^\lambda|^2=1$ and the fact that $|R_\lambda(k,p)|=1$. Let us consider values of $y$ such that $|y|\ll W$, which means that we are keeping very close to the edge at $y=0$. In this case each term in the sum over $m$ is not much different from the next one and we can exchange the sum over $m$ for an integral with respect to $p$. More precisely, to evaluate the sum over $m$ we can use the Euler--Maclaurin formula, which will give us an expansion in powers of $1/W$ and the leading-order term is obtained by simply exchanging the sum over $m$ for an integral over $m$. This in turn can be exchanged for an integral over $p$ via $p=\pi m/W$. Taking into account Eq. \eqref{eq:norm}, this will lead to
\be
&&Q_{s}(y)=4e^2E\sum_\lambda\int\limits_{p>0}\frac{dp}{2\pi}\int\limits_{S(k)=1}\frac{dk}{2\pi}\Big(\partial_{k}f_{k,p}^\lambda\Big)\nonumber\\&&\quad\quad{}\times\Big(1+\Re \Big( [v^\lambda_{k,p}]^\dag v^\lambda_{k,-p}R_\lambda(k, p)e^{-2ip y}\Big) \Big),
\ee
 where we neglected the terms of order $O(1/W)$. Contribution of the first term in the outer round brackets in the second line to the integral is equal to zero as $f^\lambda_{k,p}=\theta(\varepsilon_{\rm F}-\varepsilon_\lambda(k,p))$ vanishes at the values of $k$ far enough from the bottom of the valley. Therefore we are left with
\be\label{eq:what_you_love}
&&Q_{s}(y)=4e^2E\sum_\lambda\int\limits_{p>0}\frac{dp}{2\pi}\int\limits_{S(k)=1}\frac{dk}{2\pi}\Big(\partial_{k}f_{k,p}^\lambda\Big)\nonumber\\&&\qquad\qquad\quad{}\times\Re\Big( [v^\lambda_{k,p}]^\dag v^\lambda_{k,-p}R_\lambda(k, p)e^{-2ip y}\Big).
\ee
 By Riemann--Lebesgue lemma, $Q_{s}(y)\to 0$ (up to terms of order $O(1/W)$ that we discarded) as $y\to \infty$. Because integration in Eq.~\eqref{eq:what_you_love} is restricted to the Fermi surface, it is clear that $Q_{s}(y)$ oscillates in space at a wavelength corresponding to double the Fermi momentum.
 We can now introduce an ``effective current'' feeding valley number accumulation at the edge
\be
&&I_s^b=\int\limits_{-\infty}^0dy\, Q_s(y)\nonumber\\&&\quad{}=-2 e^2E\sum_\lambda\int\limits_{S(k)=1}\frac{dk}{2\pi}\int\limits_{p>0}\frac{dp}{2\pi}\Big(\partial_{k}f_{k,p}^\lambda \Big)\nonumber\\&&\quad\qquad\qquad\times{}\Im \bigg[ \frac{1}{p+i0} [v^\lambda_{k,p}]^\dag v^\lambda_{k,-p}R_\lambda(k, p)\bigg].
\ee
\section{Graphene nanoribbon. Description of the stationary states}
\label{sec:states}
In this section and the next we prove analytically that in the completely gapped state the valley Hall current for a graphene nanoribbon with zig-zag edges is non-zero and quantized.  The treatment here closely follows that of Ref. \cite{nanoribbon}. Because the spin-orbit coupling in graphene is small, we will neglect it completely. In this case spin polarization is a good quantum number and its only effect is the introduction of a degeneracy factor of $2$ in all linear response formulae. We will avoid that by considering a spinless electron from now on. For a spinful electron the results given here will have to be multiplied by a factor of 2.

As promised in Section \ref{sec:setup}, the pair of coordinates $(x,y)$ will be replaced by unit cell number $l$ and combined index $(m,\sigma)$, where $m=1\dots N$ and $\sigma=A,B$. Integrals over $x'$ and $y'$ within the unit cell will be replaced as follows
\be
\int_0^1 dx'\int_0^y dy' \to -\sum_{m'=1}^m\sum_{\sigma=A,B}
~.
\ee
Furthermore, in these expressions, the current that flows between rows $m$ and $m+1$ is denoted as $j^y_{v}(m+1/2, \omega)$.

The tight binding Hamiltonian for  graphene nanoribbon with zig-zag edges in second quantized form reads 
\be
&&\hat{H}=-t\sum_{l}\bigg[\sum_{m=1}^{N}\hat{a}_l^\dag(m)\hat{b}_l(m)+\sum_{m=1}^{N-1}\hat{a}_l^\dag(m+1)\hat{b}_l(m)\nonumber\\&&\qquad\qquad\qquad\qquad{}+\sum_{m\in even}\hat{a}_l^\dag(m)\hat{b}_{l-1}(m)\nonumber\\&&\qquad\qquad\qquad\qquad{}+\sum_{m\in odd}\hat{a}_{l}^\dag(m) \hat{b}_{l+1}(m)+\mbox{h.c.}\bigg]\nonumber\\&&\qquad{}+\Delta\sum_{l}\sum_{m=1}^{N}\bigg[\hat{a}_l^\dag(m)\hat{a}_l(m)-\hat{b}_{l}^\dag(m)\hat{b}_{l}(m)\bigg],
\ee
 where $t$ is the nearest neighbour hopping parameter and $\hat{a}_l(m)$ and $\hat{b}_l(m)$ destroy an electron on sublattice $A$ or $B$, respectively, in a two-atom row $m$ of unit cell $l$. They satisfy the usual anticommutation relations
\be
\{\hat{a}_l(m), \hat{a}^\dag_{l'}(m')\}=\{\hat{b}_l(m), \hat{b}^\dag_{l'}(m')\}=\delta_{ll'}\delta_{mm'}.
\ee
We introduce the Fourier-transformed operators $\hat{\alpha}_{k}(m)$ and $\hat{\beta}_{k}(m)$ from the relations
\be\label{eq:Fourier_trans}
&&\hat{a}_l(m)=\int\frac{dk}{\sqrt{2\pi}}\hat{\alpha}_{k}(m)e^{ikx_{l,m,A}},\\
&&\hat{b}_l(m)=\int\frac{dk}{\sqrt{2\pi}}\hat{\beta}_{k}(m)e^{ikx_{l,m,B}},
\ee
 where $x_{l,m,A}$ and $x_{l,m,B}$ are the positions of the atoms of the $A$ and $B$ sublattices in the direction along the ribbon and $k$ takes values in the interval $(-\pi, \pi]$. The Fourier-transformed creation and annihilation operators satisfy the anticommutation relations 
\be
\{\hat{\alpha}_k(m),\hat{\alpha}^\dag_{k'}(m')\}&=&\{\hat{\beta}_k(m),\hat{\beta}^\dag_{k'}(m')\}\nonumber\\&=&\delta(k-k')\delta_{mm'}.
\ee
In terms of Fourier transformed creation and annihilation operators, the Hamiltonian reads
\be
\hat{H}=&&\int dk\bigg[-t\sum_{m=1}^{N}\hat{\alpha}_{k}^\dag(m)\hat{\beta}_k(m)g_k+\mbox{h.c.}\nonumber\\&&-t\sum_{m=1}^{N-1}\hat{\alpha}_{k}^\dag(m+1)\hat{\beta}_k(m)+\mbox{h.c.}\nonumber\\&&{}+\Delta\sum_{m=1}^{N}\Big(\hat{\alpha}_k^\dag(m)\hat{\alpha}_k(m)-\hat{\beta}_k^\dag(m)\hat{\beta}_k(m)\Big)\bigg],
\ee
 where $g_k=2\cos(k/2)$. From now on, we will set $t=1$. The one-particle eigenstates of the Hamiltonian are Bloch states $\psi_{k,n}(l,m,\sigma)=u_{k,n}(m,\sigma)\exp{(ikx_{l,m,\sigma})}/\sqrt{2\pi}$. As is customary, we will combine the amplitudes $u_{k,n}(m,\sigma)$ for $\sigma=A, B$ into a sublattice pseudospinor $u_{k,n}(m)=(u_{k,n}(m,A),\; u_{k,n}(m,B))^t$. For bulk states, this takes the form
\be \label{eq:eigenstates}
u_{kps}(m)=\mathcal{N}_{kps}\begin{pmatrix}
(\varepsilon_{kps}+\Delta)\sin \big[p(N+1-m)\big]\\
(-1)^j\sqrt{\varepsilon_{kps}^2-\Delta^2}\sin (p m)
\end{pmatrix},\nonumber\\
\ee
 where $\varepsilon_{kps}=s\sqrt{\Delta^2+g_k^2+2g_k\cos (p)+1}$ and the role of the band index $n$ is played by the combination $ps$, with  $s=\pm$ and $p\in (0,\pi)$ the solution of the equation
\be\label{eq:eigeneq}
pN+\arccos \left(\frac{1+g_k \cos p}{\sqrt{g_k^2+2g_k\cos p+1}}\right)=\pi j.
\ee
In Eqs.~\eqref{eq:eigenstates}--\eqref{eq:eigeneq}, $j=1,2,\dots, N$ for $g_k>N/(N+1)$ and $j=1,2,\dots, N-1$ for $g_k<N/(N+1)$. 

In Eq.~\eqref{eq:eigenstates} the normalization constant equals
\be
&&\mathcal{N}_{kps}=\Bigg[(\varepsilon_{kps}+\Delta)\varepsilon_{kps}\nonumber\\&&\qquad\qquad{}\times\Bigg(N-\frac{\sin (p N)\cos \big[p(N+1)\big]}{\sin (p)}\Bigg)\Bigg]^{-1/2}.\nonumber\\
\ee
We note that, for $g_k<N/(N+1)$, apart from the states described by Eqs.~\eqref{eq:eigenstates}--\eqref{eq:eigeneq}, there is also an edge state
\be
&&u_{kes}\nonumber\\&&{}=\mathcal{N}_{kes}(-1)^m\begin{pmatrix}
(\varepsilon_{kes}+\Delta)\sinh\big[(N+1-m){\eta_k}\big]\\
-\sqrt{\varepsilon_{kes}^2-\Delta^2}\sinh ({\eta_k} m)
\end{pmatrix},\nonumber\\
\ee
 where $\varepsilon_{kes}=s\sqrt{\Delta^2+g_k^2-2g_k\cosh ({\eta_k})+1}$, with $s=\pm$, and ${\eta_k}>0$ is the solution of the equation
\be\label{eq:edge_eigeneq}
e^{2{\eta_k}(N+1)}=\frac{g_k-e^{\eta_k}}{g_k-e^{-{\eta_k}}}.
\ee
 The role of the band index $n$ here is played by the combination $es$, where $e$ stands for `edge' and $s$ is described above. The normalization factor equals
\be
&&\mathcal{N}_{kes}=\bigg[\varepsilon_{kes}(\varepsilon_{kes}+\Delta)\nonumber\\&&\quad\qquad{}\times\left(\frac{\cosh\big[(N+1){\eta_k}\big] \sinh (N{\eta_k})}{\sinh ({\eta_k})}-N\right)\bigg]^{-1/2}.\nonumber\\
\ee

\section{Current is quantized for the totally gapped graphene nanoribbon}
Let us calculate the valley Hall current for the totally gapped graphene nanoribbon with zig-zag edges in the limit of vanishing frequency. Using Eq. \eqref{eq:zero_freq_curr} (and neglecting spin) we obtain
\be\label{eq:cur_resp}
 j_{v}^y(m+1/2, 0)&=&e^2E\sum_{n}\int \frac{dk}{2\pi}f_{k,n}\,S(k)
\nonumber\\
&\times&
\sum_{m'=1}^{m}\partial_k(u_{k,n}^\dag(m')u_{k,n}(m')),
\ee
 where $n$ stands for either $ps$ or $es$, as explained in Sect.~\ref{sec:states}. Note that explicit summation over $\sigma$ here is replaced by matrix multiplication of the hermitian conjugate of a pseudospinor with itself. When the chemical potential is in the gap, $f_{kps}=f_{kes}$ and equals $1$ for $s=-1$ and $0$ for $s=+1$. For the evaluation of the integral with respect to $k$ note that for any $n$ holds $u_{k,n}=u_{-k,n}$ and we can take $S(k)$ equal to $1$ for $0<k<\pi$ and $-1$ for $-\pi<k<0$. Thus keeping only the occupied states in the sum over $n$ and evaluating the integral with respect to $k$, we obtain
\be\label{eq:integrated}
 j^y_{v}(m+1/2, 0)&=&\frac{e^2E}{\pi}\sum_{m'=1}^m\bigg(\sum_{p} u_{kp-}^\dag(m')u_{kp-}(m')
\nonumber\\
&+& u_{ke-}^\dag(m')u_{ke-}(m')\bigg)\bigg|_{k=0}^{k=\pi}.
\ee
 The states $u_{k,n}$ form a complete set, therefore
\be\label{eq:completeness}
&&\sum_{p}\bigg(u_{kp+}(m')u^\dag_{kp+}(m')+u_{kp-}(m')u^\dag_{kp-}(m')\bigg)\nonumber\\&&{}+u_{ke+}(m')u^\dag_{ke+}(m')+u_{ke-}(m')u^\dag_{ke-}(m')=1_2,\nonumber\\
\ee
 where $1_2$ is the $2\times 2$ identity matrix in the sublattice space. Using the explicit form of the wavefunctions [see Eq. \eqref{eq:eigenstates}] one can find that
\be\label{eq:come_in_handy}
&&u^\dag_{kp-}(m')u_{kp-}(m')-u^\dag_{kp+}(m')u_{kp+}(m')\nonumber\\&&\quad=\frac{2\Delta}{\varepsilon_{kp-}}\cdot\frac{\sin\big[p(N+1-2m')\big]\sin\big[p(N+1)\big]}{N-\sin(pN)\cos\big[p(N+1)\big]/\sin(p)}.\nonumber\\
\ee
Using Eqs. \eqref{eq:completeness} and \eqref{eq:come_in_handy} one can rewrite the sum in the round brackets in Eq. \eqref{eq:integrated} in the form
\begin{widetext}
\be
\sum_{p}u^\dag_{kp-}(m')u_{kp-}(m')+u^\dag_{ke-}(m')u_{ke-}(m')=1&+&\frac{1}{2}\left(u^\dag_{ke-}(m')u_{ke-}(m')-u^\dag_{ke+}(m')u_{ke+}(m')\right)\nonumber\\&+&\sum_{p}\frac{\Delta}{\varepsilon_{kp-}}\cdot\frac{\sin \big[p(N+1-2m')\big]\sin \big[p(N+1)\big]}{N-\sin (pN)\cos \big[p(N+1)\big]/\sin (p)}.
\ee
Plugging this into Eq.~\eqref{eq:integrated} and taking into account the fact that there are no edge states at $k=0$ we obtain
\be\label{eq:preliminary}
 j^y_{v}(m+1/2, 0)&=&\sum_{m'=1}^m\frac{e^2E}{2\pi}\left(u^\dag_{\pi e-}(m')u_{\pi e-}(m')-u^\dag_{\pi e+}(m')u_{\pi e+}(m')\right)
\nonumber\\
&+&
\frac{e^2E}{\pi}\sum_{p}\frac{\Delta}{\varepsilon_{kp-}}\cdot\frac{\sin (p m)\sin \big[p(N-m)\big]\sin \big[p(N+1)\big]/\sin (p)}{N-\sin (pN) \cos \big[p(N+1)\big]/\sin (p)}\bigg|_{k=0}^{k=\pi}.
\ee
\end{widetext}
 Let us estimate the sum over $p$ in Eq.~\eqref{eq:preliminary}. Since by Eq.~\eqref{eq:eigeneq}, the factors
\be
&& \frac{\sin \big[p(N+1)\big]}{\sin (p)} = \frac{(-1)^j}{\sqrt{g_k^2+2g_k\cos (p)+1}},
\\
&& \frac{\sin (pN)}{\sin (p)} = \frac{(-1)^{j-1}g_k}{\sqrt{g_k^2+2g_k\cos (p)+1}},
\ee
 remain of order $O(N^0)$ at $k=0$ or $k=\pi$, one can see that each term in the sum in the second line of Eq.~\eqref{eq:preliminary} is of order $\Delta/N$. In making this estimate we also took into account that $\varepsilon_{kp-}$ is of order one [in units of $t$] at $k=0$ and $k=\pi$. Therefore the whole sum over $p$ in Eq.~\eqref{eq:preliminary} is of order $\Delta$.  Hence the current density response equals (restoring the hopping parameter $t$)
\be\label{eq:semifinal}
&& j_v^y(m+1/2, 0) = \sum_{m'=1}^m\frac{e^2E}{2\pi}\bigg(u^\dag_{\pi e-}(m')u_{\pi e-}(m')\nonumber\\
&&\qquad\quad\quad\quad{}-
u^\dag_{\pi e+}(m')u_{\pi e+}(m')\bigg)+O(\Delta/t).
\ee
 For edge states,  as $k$ approaches $\pi$,  $g_k\to 0$ while ${\eta_k}$ approaches positive infinity as ${\eta_k} \propto -\ln(g_k)$. In this limit $\varepsilon_{kes}\to s|\Delta|$ and the probability distribution for the edge states has the form
\be\label{eq:useful_later}
u^\dag_{\pi es}(m')u_{\pi es}(m') &=& \frac{1}{2}\bigg(\Big[1+s \cdot\mbox{sign}(\Delta)\Big]\delta_{m',1}
\nonumber\\
&+&\Big[1-s\cdot\mbox{sign}(\Delta)\Big]\delta_{m',N}\bigg).
\ee
 Plugging this into Eq.~\eqref{eq:semifinal}, for $1\leq m \leq N-1$ we obtain  
\be\label{eq:curr_curr}
 j_v^y(m+1/2, 0)=-\frac{e^2E}{2\pi}\mbox{sign}(\Delta)+O(\Delta/t).
\ee
 Let us also, for future reference, quote here the result for the divergence of the current, ${\rm div}\, j_v^y(m,0) =  j_v^y(m-1/2, 0)- j_v^y(m+1/2,0)$, which follows from this equation
\be\label{eq:curr_divergence} 
{\rm div}\, j_v^y(m,0)=\frac{e^2E}{2\pi}{\rm sign}(\Delta)(\delta_{m,1}-\delta_{m,N})+O(\Delta/t).\nonumber\\
\ee

\section{Qualitative explanation of the valley Hall current profile for different boundary conditions}
Throughout this section we assume that the Fermi level lies in the conduction band and takes a value in the interval $|\Delta|<\varepsilon_{\rm F}<t$. Let us first consider the zig-zag nanoribbon. Expression for the current, Eq. \eqref{eq:cur_resp}, can be written in an alternative form as $ j_v^y(m+1/2, 0)=Ee^2(2\pi)^{-1}\big[T(m+1/2)+G(m+1/2)\big]$, where for $1\leq m \leq N-1$
\be \label{eq:def_T}
&&T(m+1/2)\nonumber\\
&&\quad{}=\sum_{n}\sum_{m'=1}^m\int dk S(k)\partial_k\Big(f_{k,n}u_{k,n}^\dag(m')u_{k,n}(m')\Big)\nonumber\\
\ee
 and
\be
&&G(m+1/2)\nonumber\\
&&\quad{}=-\sum_{n}\sum_{m'=1}^{m}\int dk S(k)(\partial_k f_{k,n})u_{k,n}^\dag(m')u_{k,n}(m')\nonumber\\
\ee
 and we set $T(1/2)=G(1/2)=T(N+1/2)=G(N+1/2)=0$. Let us consider $T$ first. Consider ${\rm div}\, T(m)=T(m-1/2)-T(m+1/2)$, which is given by the equation
\be
{\rm div}\,T(m)=-\sum_{n}\int dk S(k)\partial_k\Big(f_{k,n}u_{k,n}^\dag(m)u_{k,n}(m)\Big).\nonumber\\
\ee
 The integral with respect to $k$ can be easily calculated to produce
\be
{\rm div}\,T(m)=-2\sum_{n}f_{k,n}u_{k,n}^\dag(m)u_{k,n}(m)\Big|^{k=\pi}_{k=0},
\ee
 where a factor of $2$ appeared because the contribution of the left valley to the integral equals the contribution of the right valley, so we left only the latter and multiplied it by two. Plugging in the occupation numbers, we obtain
\be\label{eq:eval_T}
{\rm div}\,T(m)=&&{}-2\,u_{\pi e+}^\dag(m)u_{\pi e+}(m)\nonumber\\
&&{}-2\bigg(\sum_{p} u_{kp-}^\dag(m)u_{kp-}(m)\nonumber\\
&&\quad\qquad{}+ u_{ke-}^\dag(m)u_{ke-}(m)\bigg)\bigg|_{k=0}^{k=\pi}.
\ee
 The sum in the second and third lines on the right hand side of this equation (including the factor of $-2$) has already been calculated in the previous section, compare Eq. \eqref{eq:integrated}. It is given by whatever multiplies $e^2E/(2\pi)$ in Eq. \eqref{eq:curr_divergence}. The term in the first line is the contribution of the upper band of edge states, which has appeared because we raised the Fermi energy and this band became occupied. So, using  Eq. \eqref{eq:useful_later} and Eq. \eqref{eq:curr_divergence}, we  obtain (ignoring terms of order $O(\Delta/t)$)
\be\label{eq:curr_discont}
{\rm div}\,T(m)=-\delta_{m,1}-\delta_{m,N}.
\ee
 Note that ${\rm div}\, T$ changes extremely fast on the boundaries (it goes from $-1$ to zero on the scale of one inter-atomic distance) and does not change at all inside the ribbon. This can be traced back to the fact that ultimately the spatial behavior of $T(m)$ is governed by the localized edge states (see Eq. \eqref{eq:semifinal} and Eq. \eqref{eq:eval_T}).  

Consider now ${\rm div}\, G(m)$, which is given by the equation
\be
{\rm div}\,G(m)=\sum_{n}\int dk S(k)(\partial_k f_{k,n})u_{k,n}^\dag(m)u_{k,n}(m).\nonumber\\
\ee
 Let us point out that ${\rm div}\,G(m)$, as opposed to ${\rm div}\,T(m)$,  changes slowly in space, because its behavior is governed by the states on the Fermi surface. Therefore it varies significantly on the length scale defined by the inverse Fermi momentum $(\sqrt{\varepsilon_{{\rm F}}^2-\Delta^2})^{-1}\gg 1$ for $\varepsilon_{{\rm F}}$ close enough to the bottom of the conduction band. Therefore the roughest (but only the roughest) estimate of  ${\rm div}\,G(m)$ can be given by just the spatial average $\langle {\rm div}\,G(m) \rangle_{\rm sp}=(1/N)\sum_m {\rm div}\,G(m)$. This is not zero. Indeed, using  $\sum_{m} u_{k,n}^\dag(m)u_{k,n}(m)=1$ we obtain
\be
\langle {\rm div}\,G(m) \rangle_{\rm sp}=\frac{1}{N}\sum_n\int dk S(k)(\partial_k f_{k,n}),
\ee
 which is non-zero because of the left-mover--right-mover imbalance in each valley. Indeed, using $f_{k,n}=\theta(\varepsilon_{\rm F}-\varepsilon_{k,n})$ we obtain
\be 
\sum_{n}\partial_k f_{k,n}=-\sum_{i}\delta(k-k_i){\rm sign}(v_i),
\ee
 where the sum runs over all values of $k$ at which the Fermi level crosses an energy band and      $v_i$ is the group velocity of the band crossed at a point $k_i$. Because there is one more right-mover than there are left-movers in the left valley and one more left-mover than there are right-movers in the right valley, see Fig. 1(b) (the upper blue line) in the main text, we obtain
\be \label{eq:div_G}
\langle {\rm div}\,G(m) \rangle_{\rm sp}=\frac{2}{ N}.
\ee
 To get $ j^y_{v}(m+1/2,0)$ we need to integrate ({\it i.e.,} sum over $m$) Eq. \eqref{eq:curr_discont} and Eq. \eqref{eq:div_G} with the boundary conditions that the current vanishes outside the ribbon. It is not difficult to observe that Eq. \eqref{eq:curr_discont} determines the one-sided limiting values of the current on the boundaries as approached from within the ribbon and Eq. \eqref{eq:div_G} its overall slope as a function of position. It then follows that the current $ j_v^y(m+1/2, 0)$ is approximately equal to $Ee^2/(2\pi)$ at $m=1$, to $-Ee^2/(2\pi)$ at $m=N-1$ and those two values are connected roughly by a straight line with the slope $-[e^2E/(2\pi)](2/N)$.

Now let us briefly discuss the case of the nanoribbon with a bearded edge. Because there is only one band of edge states whose occupation number does not change as we raise $\varepsilon_{{\rm F}}$, we see that ${\rm div}\, T$ does not change compared to the undoped case. This means that the values of the current on the boundaries will stay roughly the same as in the undoped case. Next, because there is no left-mover--right-mover imbalance in the valleys, the overall average slope will be zero. This very rough analysis is confirmed by the numerical results, see Fig. 2 (a), (b) in the main text.

\section{Valley Hall current as function of the Fermi energy}

\begin{figure}[b]
\includegraphics[scale=0.6]{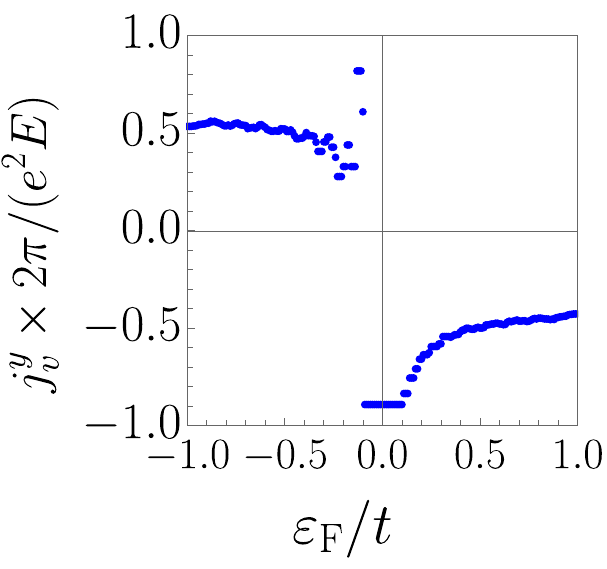}
\caption{\label{fig:curr_curr} Valley Hall current as a function of the Fermi energy in graphene nanoribbon with one edge zig-zag and the other bearded. The value is taken in the middle of the ribbon, the width $N=100$, $\Delta=0.1t$.}
\end{figure}

In this section we provide numerical results for the valley Hall current as a function of the Fermi energy in graphene nanoribbon with one edge zigzag and the other bearded, see Fig. \ref{fig:curr_curr}. In an infinite system this is predicted to be fixed and quantized when the chemical potential is in the band gap and to go down to zero as $-\Delta/|\varepsilon_{\rm F}|$ (in units of $e^2E/(2\pi)$) for $|\varepsilon_{\rm F}|>|\Delta|$. For the nanoribbon the behavior of the curve is a bit different, see Fig. \ref{fig:curr_curr}. It is indeed fixed and quantized (up to corrections of order $O(\Delta/t)$) when the Fermi energy is in the gap but outside the gap it does not conform to the $\Delta/|\varepsilon_{\rm F}|$ law and settles on a value of around $\pm 1/2$ for high enough hole or electron doping.

At $\varepsilon_{\rm F}=-|\Delta|$ the valley Hall current has a discontinuity due to all the edge states suddenly changing their occupation numbers. The oscillations below  $\varepsilon_{\rm F}=-|\Delta|$ are a finite size effect due to small values of the Fermi momentum.